\documentclass[preprint,5p,twocolumn,sort&compress]{elsarticle}
\usepackage[dvipsnames,svgnames]{xcolor}
\usepackage{lineno,hyperref}
\usepackage{amsmath}
\modulolinenumbers[1]

\journal{Journal of \LaTeX\ Templates}

\begin{document}

\begin{frontmatter}

\title{Large mass single electron resolution detector for dark matter and neutrino elastic interaction searches}

\author[niseraddress]{V. Iyer}
\cortext[mycorrespondingauthor]{Corresponding author}
\ead{vijayiyer@niser.ac.in}
\author[tamuaddress]{N. Mirabolfathi}
\author[tamuaddress]{G. Agnolet}
\author[tamuaddress]{H. Chen}
\author[tamuaddress]{A. Jastram}
\author[stanfaddress]{F. Kadribasic}
\author[niseraddress]{V. K. S. Kashyap}
\author[tamuaddress]{A. Kubik}
\author[tamuaddress]{M. Lee}
\author[tamuaddress]{R. Mahapatra}
\author[niseraddress]{B. Mohanty}
\author[tamuaddress]{H. Neog}
\author[tamuaddress]{M. Platt}



\address[niseraddress]{School of Physical Sciences, National Institute of Science Education and Research, HBNI, Jatni 752050, India}
\address[tamuaddress]{Department of Physics and Astronomy, and the Mitchell Institute for Fundamental Physics and Astronomy, Texas
A\&M University, College Station, TX 77843, USA}
\address[stanfaddress]{Department of Physics, Stanford University, Stanford, CA 94305, USA}

\begin{abstract}
Large mass single electron resolution solid state detectors are desirable to search for low mass dark matter candidates and to measure coherent elastic neutrino nucleus scattering (CE$\nu$NS). Here, we present results from a novel 100 g phonon-mediated Si detector with a new interface architecture. This detector gives a baseline resolution of $\sim 1  e^{-}/h^{+}$ pair and a leakage current on the order of $10^{-16}$ A. This was achieved by removing the direct electrical contact between the Si crystal and the metallic electrode, and by increasing the phonon absorption efficiency of the sensors.  The phonon signal amplification in the detector shows a linear increase while the signal to noise ratio improves with bias voltage, up to 240 V. This feature enables the detector to operate at a low energy threshold which is beneficial for dark matter and CE$\nu$NS like searches. 

\end{abstract}


\end{frontmatter}


\section{\label{sec:sec1}Motivation for large mass low threshold detectors for rare event searches}

There is substantial evidence that Dark Matter (DM) is the main ingredient of the matter content in the Universe. The increasing number of null results from GeV to TeV scale searches has increased interest to search for low mass DM particles in the sub-GeV scale \cite{SNOWMASS:2013}. Several theories in support of low mass DM such as Asymmetric DM \cite{ADM}, or DM-Standard Model (SM) coupling through dark photons \cite{DarkPhotons} are also shifting the focus of experimentalists towards low mass DM searches. For  direct  searches  that  seek  detection  via  DM  elastic  scattering off the atomic nuclei,  the ratio of DM mass ($M_{DM}$) to target nucleus mass reduces the energy expected ($E_{th}$) from DM interactions for smaller DM masses, as $E_{th} \propto M_{DM}^{2}$. This together with the weak scale rate expected from DM interactions necessitates detectors with relatively large masses ($\sim$kg scale) and very low thresholds ($\sim$eV scale).

The standard model predicts that neutrinos can scatter elastically off atomic nuclei through a coherent process in which the cross section of interaction scales as $A^2$, where A is the mass number of the target material \cite{Freedman}. Coherent elastic neutrino-nucleus scattering (CE$\nu$NS) forms an irreducible background for DM searches. CE$\nu$NS has been experimentally observed for neutrinos with $E_{\nu}$ $\sim$ 20 MeV \cite{COHERENT}, but has not been measured for lower energy neutrinos from reactors. To measure these low energy neutrino-nucleus scattering, detectors with low recoil energy thresholds ($\sim 100$ eV ) are desirable. Because the quantum excitation in solid state ionization mediated detectors are very small ($\sim1$ eV), these detectors would have sufficient sensitivity to detect  DM interacting directly with the valence electrons or indirectly through nuclear recoil.

Due  to  technical  challenges  inherent  in  high-impedance charge-amplifier readouts, direct ionization measurement signal to noise (S/N) ratio is limited by the readout front-end input capacitance. Although high resolution CCD based ionization readout offers single electron resolution measurements, the technique is limited to very small mass detector ($<$ 10 g) modules. A  new measurement technique has been developed by the Cryogenic Dark Matter Search (CDMS) experiment wherein ionization is indirectly measured via the phonons generated by the carrier drift in the crystal, known as the Neganov-Trofimov-Luke (NTL)\cite{Neganov, Luke} effect. This results in a phonon signal that is proportional to both the applied potential across the crystal and the number of electron-hole pairs created after an interaction. The phonon signals are measured using low impedance superconducting transition edge sensors\cite{QET:1995} that are photolithographically deposited on the surface of the crystal. The low impedance phonon readout is inherently more immune to the environmental noise and thus offers a far better S/N than direct ionization readouts.  Moreover, as the phonon readout is in general independent of the high voltage bias, we can expect a S/N that grows linearly with the bias.  However, above a certain biasing threshold, stochastic carrier leakage into the crystal amplified by the NTL process appears as irreducible noise. This limits the ultimate sensitivity of NTL-phonon-assisted readout technology. The early onset of this spurious leakage current observed with CDMSlite detectors has been the limiting factor for the sensitivity of the experiment.  Here we present our recent progress in reducing this leakage current and in turn the energy resolution, using a new detector interface geometry and the detector sensitivity gains thereof.

\subsection{\label{sec:sec12}Detector fabrication}

\begin{figure}[t!]
\centering
\includegraphics[width=0.65\linewidth]{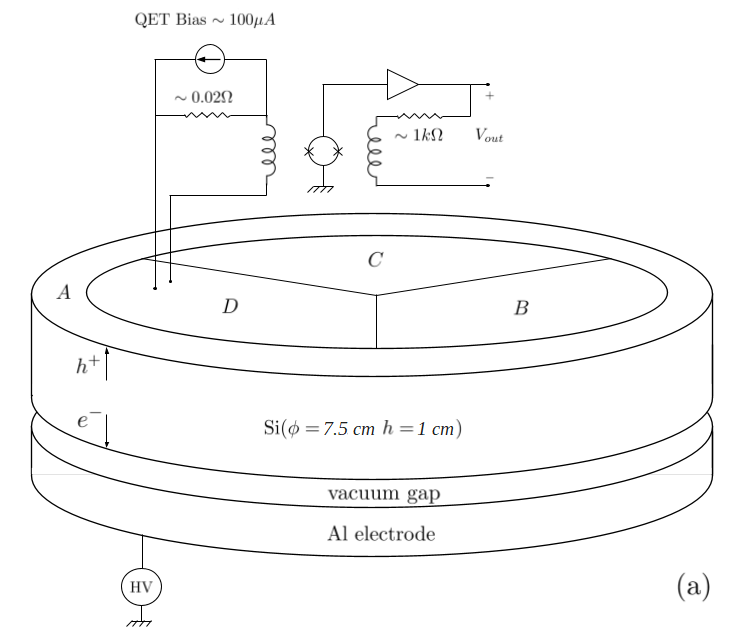}
\includegraphics[width=0.6\linewidth]{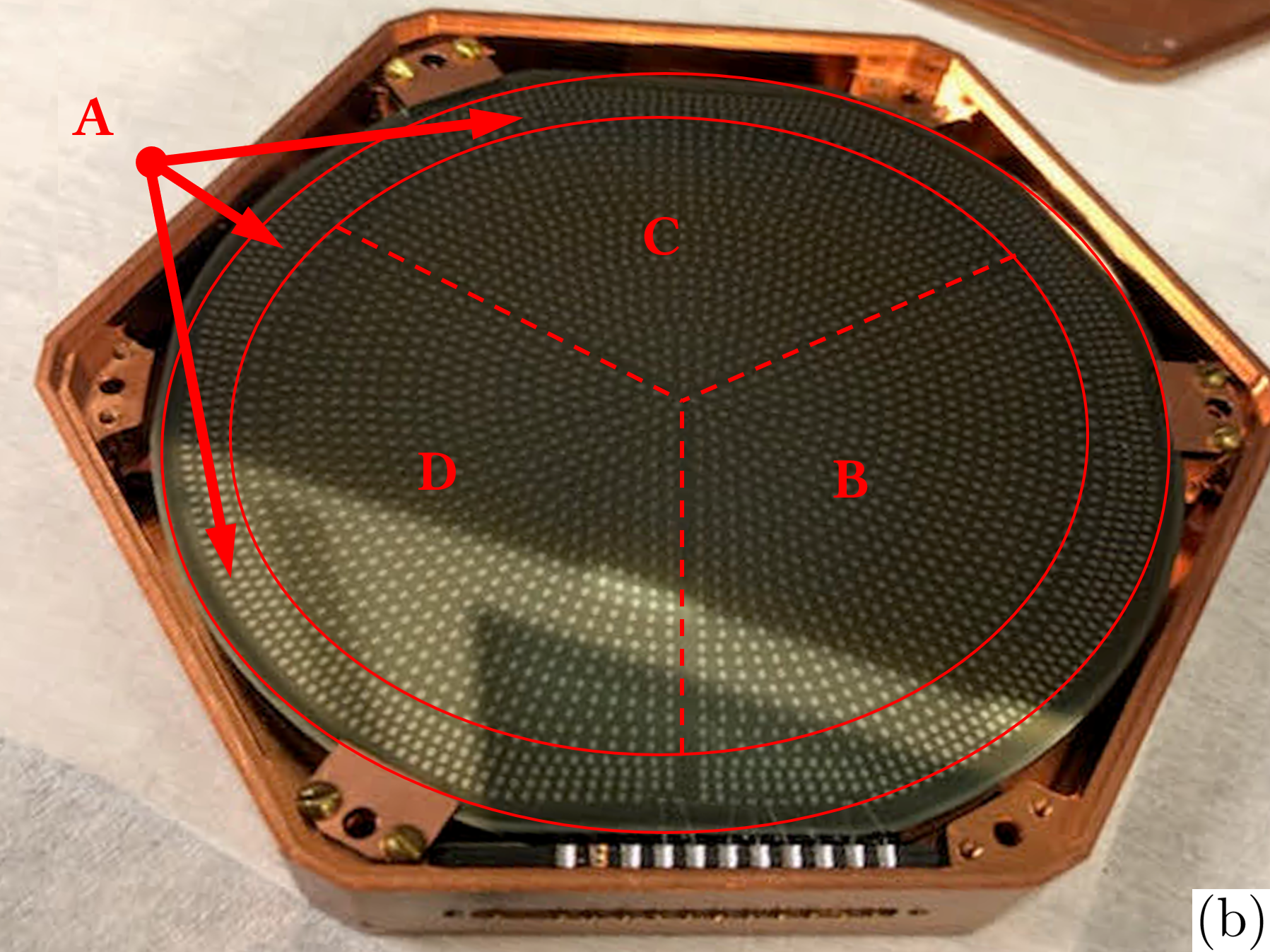}
\caption{\label{fig:Fig1} (a) Detector readout and bias circuit schematic. A 100 g silicon crystal (diameter=7.5 cm and thickness=1 cm) has one face covered with transition edge sensors to measure both primary phonons and phonons generated as the carriers drift through the crystal. The phonon sensors are held at the ground potential and the crystal is biased through a vacuum gap using an aluminum electrode. The phonon signal is read out using a SQUID front-end amplifier. (b) A picture of the detector. The phonon sensors consist of $\sim$1000 TES sensors connected in a parallel configuration. The TES sensors are divided into 4 groups of $\sim$250 sensors and independently read out forming 4 channels (A, B, C, and D). This configuration allows reconstruction of interaction locations based on the relative amplitudes and time delays between independent channels.}
\end{figure}

Our previous studies have shown that the CDMSlite's early onset of carrier leakage is predominantly due to the particular electrode-crystal interface \cite{BasuThakur}. CDMS phonon readout consists of arrays of transition edge sensors (TES) that uniformly cover both faces of the disk shape semiconductor (Ge or Si). The sensors are further grouped into 4 channels (A, B, C, and D) with a geometry shown in Fig. \ref{fig:Fig1} that are independently read out using superconducting quantum interference device (SQUID) based front-end amplifiers. This division of phonon arrays allows event localization in the detector that we will discuss later in this paper.

In an attempt to reduce carrier leakage through the metal-crystal interface, we have developed a method whereby one face of the crystal is biased through a vacuum gap of $\sim600$ $\mu$m as shown in Fig. \ref{fig:Fig1} (a). In an earlier work with a germanium detector, we had demonstrated a significant reduction of the leakage current of the detector and the best resolution obtained thereof by removing the direct electrical contact between a Ge crystal and the Al electrode \cite{Mirabolfathi}. Here, we have studied the same effect in a Si substrate of similar size. Furthermore, on the phonon readout side of the detector (that also provides the ground potential), we have improved the interface by removing the amorphous-Si layer and establishing a direct Si/electrode Schottky contact. Amorphous-Si layers were historically introduced in the CDMS contact architecture to mitigate the problem of a dead layer \cite{CDMSII, DeadLayer}. Due to the large bias fields present for the NTL-phonon assisted detectors and the shallower dead layer depths thereof, the advantage of those interface layers becomes almost obsolete. Eliminating the amorphous-Si will also improve the athermal phonon (\textit{i.e.} energy) absorption efficiency by the phonon sensors resulting in an enhanced S/N which is critical for single electron resolution detectors.  

\section{\label{sec:sec2}Experimental setup and data sets}

The detector assembly is mounted inside a BlueFors LD400 pulse-tube based $^{3}$He--$^{4}$He dilution refrigerator. Photons of energy 1.9 eV (640 nm) are transported from a room temperature pulsed laser (6-60 ns pulse width) via a single mode optical fiber to the cold volume of the refrigerator where the detector resides. The cold end of the fiber directly projects onto the polished surface of the detector via a gap of $\sim1$ mm under channel B of the detector. To calibrate the energy scale, an $^{55}$Fe source is placed under channel D, on the bare and polished surface of the detector. The detector is mounted with CDMS SQUID front-end amplifiers. The laser is pulsed synchronously with the data acquisition setup to mitigate the influence of the high rates expected from the ambient radioactivity. The response of the detector was measured over a range of high voltage (HV) biases to study the NTL gain in phonon signal and noise to assess the S/N in this new detector configuration. 

Due to the vacuum gap between the Si crystal and the Al electrode, the effective bias voltage across the crystal (HV$_{effective}$) is less than the applied HV across the circuit (HV$_{applied}$). The ratio between HV$_\mathrm{effective}$ and HV$_{applied}$ is determined to be $\sim 0.57$ using the NTL gain factor for the $^{55}$Fe 5.89 keV line between 0 V and 20 V bias. If we consider the silicon and vacuum gap to act as two capacitors connected in series, then using the voltage divider formula the gap is estimated to be $\sim 600$  $\mu$m using the following calculations:

    \begin{align}
    \label{Eq:VoltageDivider}
        HV_\mathrm{effective}&=\frac{HV_\mathrm{applied}\times C_\mathrm{vac}}{C_\mathrm{Si}+C_\mathrm{vac}} = \frac{HV_\mathrm{applied}\dfrac{A_\mathrm{vac}\epsilon_\mathrm{vac}}{d_\mathrm{vac}}}{\dfrac{A_\mathrm{Si}\epsilon_\mathrm{Si}}{d_\mathrm{Si}}+\dfrac{A_\mathrm{vac}\epsilon_\mathrm{vac}}{d_\mathrm{vac}}}
    \end{align}

    where, $C_{vac}$ and $C_{Si}$ are the capacitances of the Silicon and vacuum media connected in series. Considering them as parallel plate capacitors, the capacitance can be expressed as $C=A\epsilon/d$ where $A$ is the area of the capacitor plates, $\epsilon$ is the permittivity of the medium, and $d$ is the distance between the two plates of a capacitor. The subscripts `Si' and `vac' represent the silicon and vacuum mediums respectively. For our case, $A_\mathrm{vac}=A_\mathrm{Si}$. 
   
    Eq. \ref{Eq:VoltageDivider} simplifies to:
    
    \begin{equation}\label{Eq:dvac}
        d_\mathrm{{vac}} = \frac{\epsilon_\mathrm{vac}}{\epsilon_\mathrm{Si}}d_\mathrm{{Si}}\left(\frac{HV_\mathrm{applied}}{HV_\mathrm{effective}}-1\right)
    \end{equation}
   
    If we substitute the values in Table~\ref{Table:dvacPars} in Eq.~\ref{Eq:dvac}, the vacuum gap is estimated to be $\sim 600$ $\mu$m.  
     
    \begin{table}[h] 
        \centering
        \caption{The values of the quantities used in Eq.~\ref{Eq:dvac}}
        \label{Table:dvacPars}
        \begin{tabular}{|c|c|c|c|}
        \hline
             $\epsilon_\mathrm{Si}$ & $\epsilon_\mathrm{vac}$ & $d_\mathrm{Si}$ (mm) & $HV_\mathrm{applied}/HV_\mathrm{effective}$ \\\hline
              12 & 1 & 10 & 1.75\\\hline
        \end{tabular}

    \end{table}

The laser pulse energy is calibrated against the $^{55}$Fe photons using all four channels of the detector at 0 V bias. An energy calibration is performed at 0 V in order to avoid systematic differences associated with the ionization quantum yield between 1.9 eV (laser) photons and 5.89 keV ($^{55}$Fe) photons. Because the $e^{-}/h^{+}$ pairs produced by the recoils recombine within the crystal, we expect that at 0 V the phonon signals are proportional to the recoil energy and independent of the quantum yield. We expect two distinct peaks in the total phonon energy summed over all channels, one from the laser photons that is collimated under channel B and another from the $^{55}$Fe 5.89 keV photons collimated under channel D. 

We construct two variables to estimate the location of the sources on the detector in a system similar to the X-Y Cartesian co-ordinates referred to as the X partition and Y partition. These variables are defined as:

\small{
\begin{flalign}\label{eq:Eq6}
\mbox{X partition}=&\displaystyle{\frac{\cos30^{\circ}\cdot pD + \cos150^{\circ}\cdot pB + \cos270^{\circ}\cdot pC}{pB+pC+pD}}\\
\mbox{Y partition}=&\displaystyle{\frac{\sin30^{\circ}\cdot pD + \sin150^{\circ}\cdot pB + \sin270^{\circ}\cdot pC}{pB+pC+pD}}\label{eq:Eq5}
\end{flalign}
}

where, $pB, pC$ and $pD$ are the phonon amplitudes of the event in the channels B, C, and D respectively. 

\begin{figure}[h!]
\centering
\includegraphics[width=\linewidth]{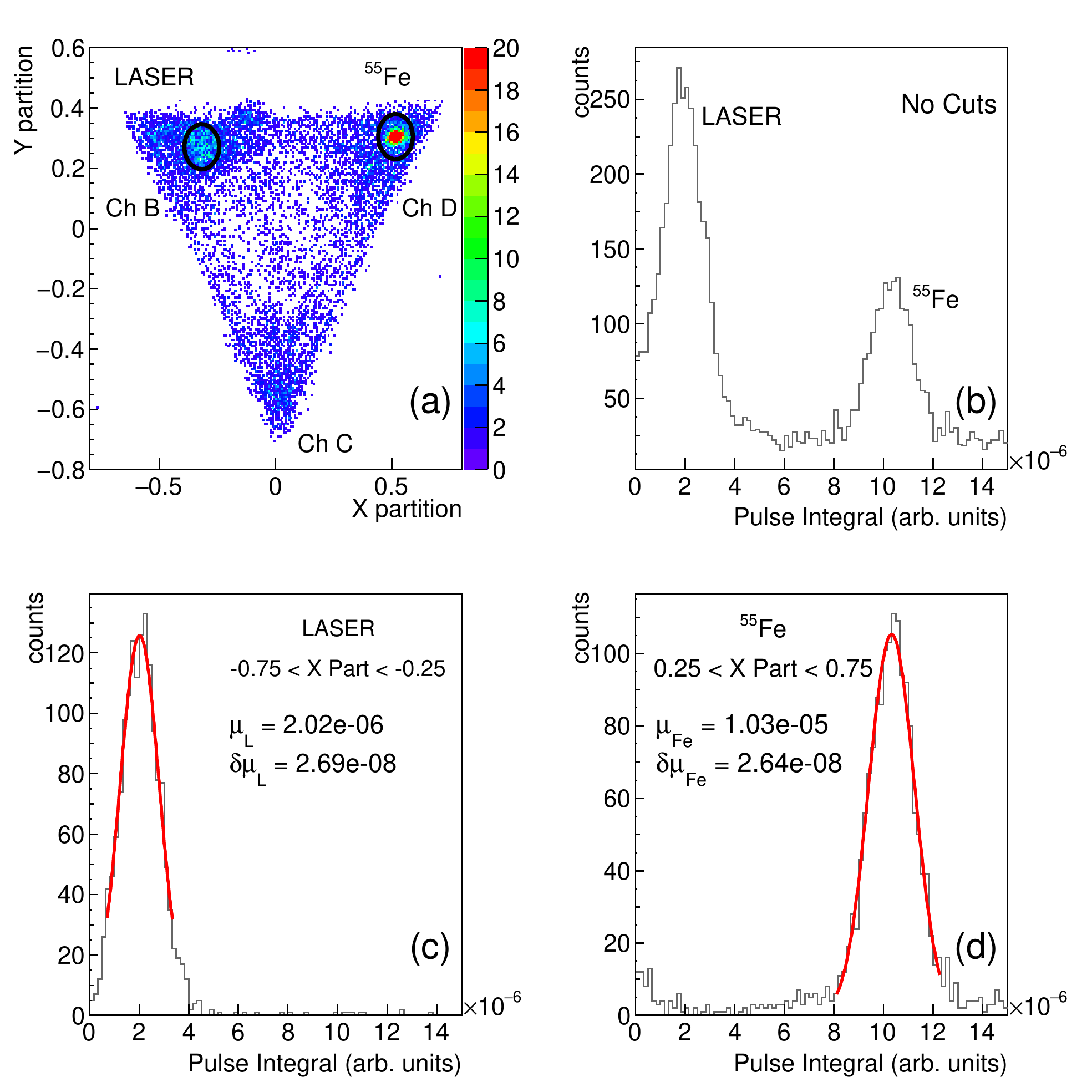}
\caption{\label{fig:Fig3} (a) A scatter plot of the Y partition variable against the X partition variable showing the location of laser events on channel B and $^{55}$Fe events on channel D. Here the colour palettes represent the number of events in each bin. Events originating in channel A would comprise of surface events and are not considered by definition in the partition variables defined by Eq.~\ref{eq:Eq6} and Eq.~\ref{eq:Eq5}. (b) the pulse integral distribution showing the laser and the $^{55}$Fe peaks at 0 V. (c) The pulse integral distribution at 0 V after applying the selection cut on the partition variables to isolate the laser peak. (d) The pulse integral distribution at 0 V after applying the selection cut on the partition variables to isolate the $^{55}$Fe peak. Panels (c) and (d) also show the Gaussian function fits to the peaks (red curves) along with the corresponding means ($\mu$) and the errors on the mean.}

\end{figure}

Fig. \ref{fig:Fig3} (a) shows the X-Y partition scatter plot. In Fig. \ref{fig:Fig3} (b), we show the total phonon energy distribution summed in arbitrary units over all channels. As expected, we observe two peaks; one from the laser source and one from the $^{55}$Fe source. By applying event selection cuts on the partition variables, we have isolated the peaks from the laser and $^{55}$Fe sources. A Gaussian function is fitted to the laser and $^{55}$Fe peaks as shown in Fig. \ref{fig:Fig3} (c) and Fig. \ref{fig:Fig3} (d). 
The average phonon energy of the laser with an unbiased detector is calculated using the ratio of the laser peak to the $^{55}$Fe peak and its value is found to be $\sim$ 1150$\pm 50$ eV.

As the laser is collimated under channel B and the majority of the phonons are absorbed in that channel, henceforth we will discuss the results only from channel B. Laser data and randomly triggered noise data were taken from 0 V to 320 V at intervals of 20 V. We expect the signal to scale linearly with voltage due to the NTL gain. The very large NTL gain at higher voltages may warm the TES sensors from the transition region to the normal state, creating a nonlinear readout. To overcome this effect, the laser intensity beyond 100 V is reduced from $\sim$ 1150 eV to $\sim$ 380 eV. Two sets of data were taken at 100 V, one at each laser intensity. The mean phonon energies in arb. units for the laser intensity at 1150 eV and 380 eV at 100 V are $5.15 \times 10^{-6}$ and $1.81 \times 10^{-6}$ respectively. The ratio of the mean phonon energies for the two laser intensities is 2.85 which is used as a scaling factor for all voltages above 100 V.

\begin{figure*}
\centering
\includegraphics[width=\linewidth]{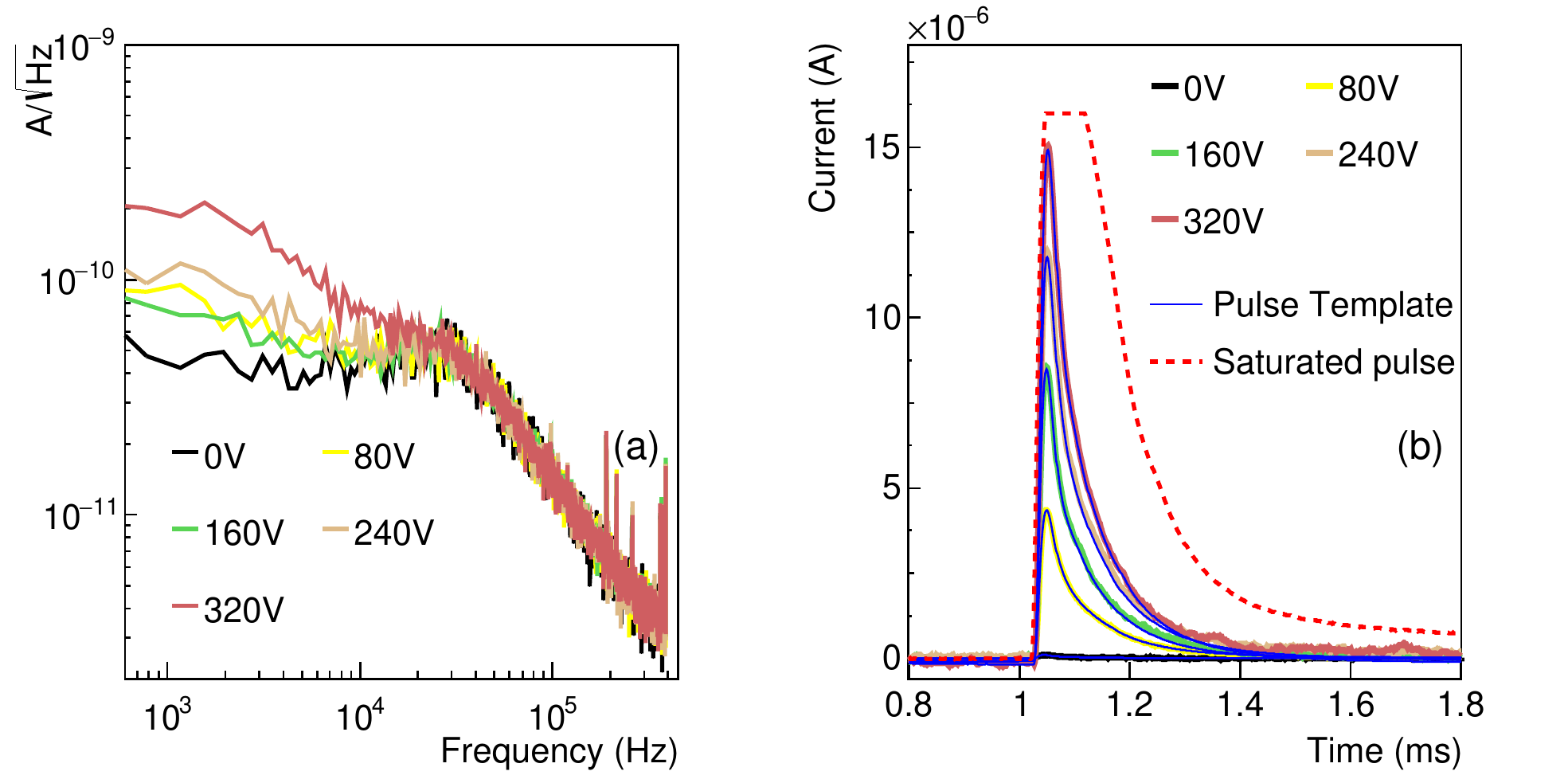}
\caption{\label{fig:Fig2} (a) The (TES current amplitude)/$\sqrt{\text{Hz}}$ plotted as a function of frequency for different voltages from 0 V to 320 V. These distributions are the phonon noise PSDs. (b) The phonon raw pulses (solid line) as current signals in the TES as a function of bias voltages. The pulses below 100 V were at a laser intensity of $\sim$ 1150 eV. The TES become saturated (dashed line) for voltages above 150 V and thus the laser intensity was reduced to $\sim$ 380 eV. The pulses above 100 V in the figure have been scaled to data below 100 V. The pulse template fit to the pulses using the OF method is also shown.}
\end{figure*}

\section{\label{sec:sec3}Data analysis and results}
To understand the phonon noise performance, we first select 10\% of the total noise events with the lowest baseline standard deviation. This is done to remove pile-ups or pulses from the randomly triggered noise data set. We look at the average power spectral density (PSD) of the noise data at each voltage. 

The noise PSD comparison across all voltages is shown in Fig. \ref{fig:Fig2}(a). The noise at low frequency ($\sim 10^{3}$ Hz) starts to increase at high voltages ($\geq240$ V) indicating a dominance of leakage current in the detector.

\begin{figure}[h!]
\centering
\includegraphics[width=0.6\linewidth]{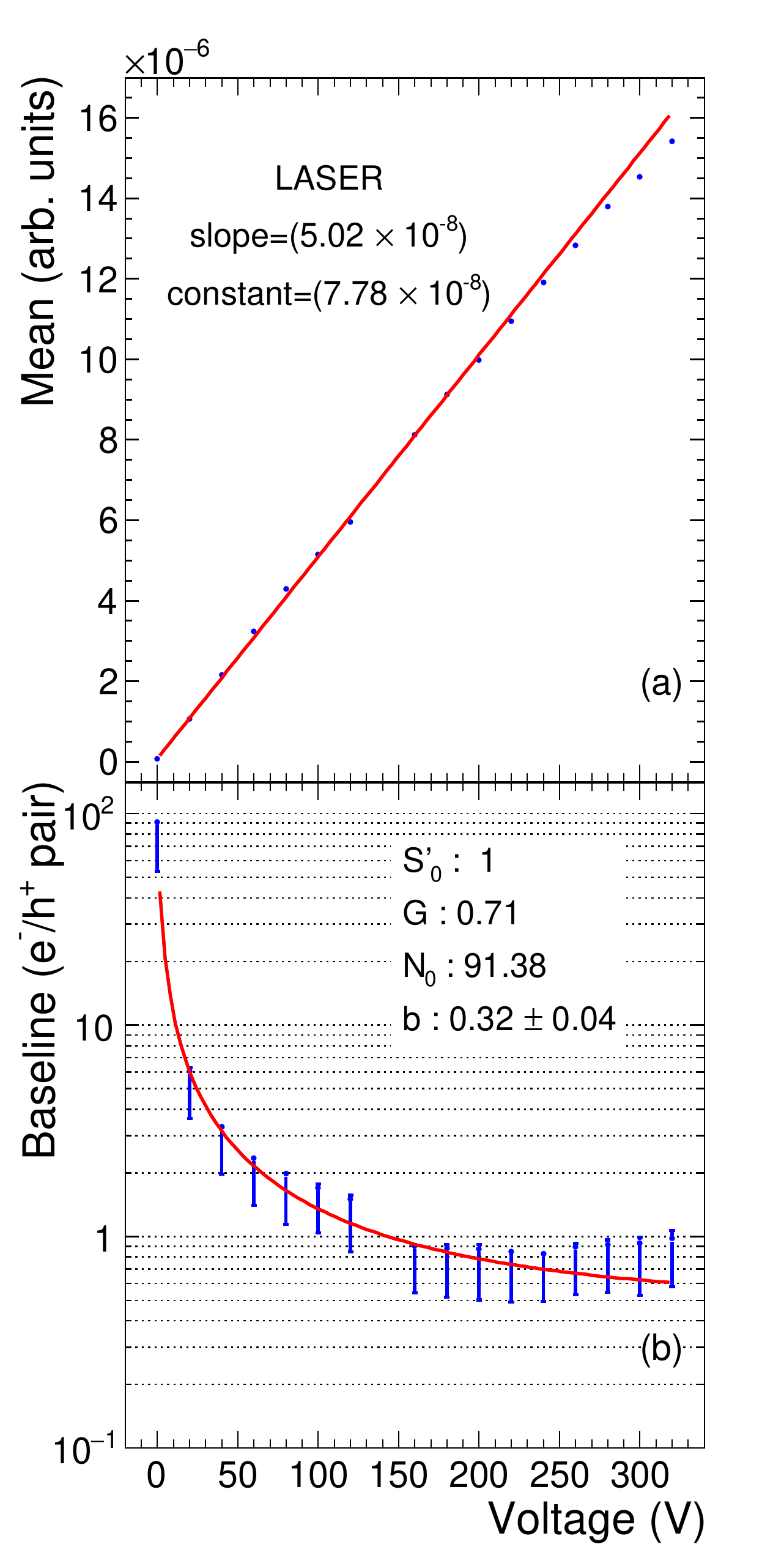}
\caption{\label{fig:Fig4} (a) The mean of the laser energies in arbitrary units obtained at different voltages from 0 V to 320 V . The red line is a straight line fit to the data points. Fit parameters are also shown. Statistical errors are small and on the order of the marker size, (b) Baseline resolution in units of $e^{-}/h^{+}$ pair units as a function of voltage. The data points are fit to a functional form which is the ratio of Eqs. \ref{eq:Eq1} and \ref{eq:Eq1a} to determine the shot noise. We found that the noise data at 140 V had been acquired during a high environmental noise period. Hence the data point is excluded from the analysis.}
\end{figure}

The amplitudes of the laser signals are obtained using the optimal filter method \cite{Golwala}. The optimal filter method fits a pulse template to each of the raw laser traces over the frequency domain to determine the laser pulse amplitude which we refer to as the OF amplitude. The pulse template is made by averaging over some select good pulses. Fig. \ref{fig:Fig2}(b) shows the pulse template fit over some laser pulses for different voltages. The good pulses are defined by putting restrictions on pulse characterizing variables such as rise time, fall time, full width half maximum, peak current amplitude, baseline standard deviation and minimum current amplitude of the trace. The OF amplitude is therefore a direct measure of the phonon energy in arbitrary energy units. We obtain the mean of the OF amplitude distribution for the laser data set at each voltage. The phonon signal amplification is expected to increase linearly with voltage due to the NTL effect \cite{Neganov, Luke}. This linear gain in amplification is observed in Fig. \ref{fig:Fig4}(a) to follow this trend up to 240 V. Above 240 V, despite the gain in phonon amplitude, the noise in the detector starts increasing more rapidly than the phonon signal thereby degrading the overall S/N.

The energy of this laser at 0 V was determined in Sec. \ref{sec:sec2} as $1150\pm50$ eV. If we consider the average energy to create an $e^{-}/h^{+}$ pair in Si with low energy photons to be equal to the band gap energy in Si ($\sim 1.12$ eV), then a laser pulse creates $\approx 1030$ $e^{-}/h^{+}$pairs. However, given the energy of the laser photons (1.9 eV) is slightly below twice the Si gap and if we assume that only 1 $e^{-}/h^{+}$ is created per photon absorption, then the quantum yield at 0 V would turn out to be $1150/1.9\approx608$. Our estimate of detector resolution depends linearly on our assumption about the quantum yield for 1.9 eV laser photons. We add this as a systematic uncertainty in this work.
The noise OF amplitude distribution can be converted to $e^{-}/h^{+}$ pair scale using the calibration factor $1150/(1.12\cdot S_n$) where $S_n$ is the mean of the distribution of the OF amplitudes and $n$ runs from 0 to 320 V in steps of 20 V as shown in Fig. \ref{fig:Fig4}(a). The noise distributions in $e^{-}/h^{+}$ pair units for each voltage are fitted to Gaussians. The sigma of each Gaussian is taken to be the baseline resolution of the detector at that voltage. Fig. \ref{fig:Fig4} (b) shows the baseline resolution as a function of voltage.

To obtain the S/N ratio and relate the observed behaviour  of the baseline resolutions as a function of the voltage, the functional forms for the noise (N) and signal (S) are taken to be

\begin{eqnarray}
\label{eq:Eq1}
N&=& \sqrt{N_0^2+(Vb)^2},\\
\label{eq:Eq1a}
S&=& S_0+S_0qVG/\epsilon,
\end{eqnarray}
where $N_0$ is the noise at 0 V, $V$ is the applied bias voltage, $b$ is the noise associated with the leakage current, $S_0$ is the mean laser amplitude at 0 V, $q$ is the charge of an electron, $G$ is a dimensionless quantity to include the effect of the vacuum gap (the effect of the voltage drop in series with the detector), and $\epsilon$ is the average energy required to create an $e^{-}/h^{+}$ pair in Si.

We expect the $S/N$ vs HV to be divided into 3 regions: (i) In the first region the dominant noise components are from thermal noise associated with the sensors, bias circuit, SQUIDS and electronics which are not dependent on the bias voltage. We expect the $S/N$ to improve linearly in this region, (ii) In the second region the noise associated with the stochastic leakage current dominates the $N$ and thus both noise and the signal increase linearly with voltage so that the $S/N$ becomes independent of bias, and  (iii) In the third region, the leakage current increases with the bias causing the $S/N$ ratio to decrease with bias. We expect the baseline resolution to degrade with voltage in this region. In the Fig. \ref{fig:Fig4} (b), we see that the resolution improves up to 120 V, then plateaus, and beyond 240 V the S/N shows signs of degradation. This behaviour is consistent with our expectations. Fig. \ref{fig:Fig4} (b) shows the $S/N$ fit using all the available baseline resolution data points  even though the fit does not represent the third region. The shot noise ($b$) is generally a function of bias voltage and its functional form in the third region is not well known. Hence the fit does not accurately represent the S/N behaviour in the third region. We obtain the lowest baseline resolution of $0.83^{+0.03}_{-0.34}$ $e^{-}/h^{+}$ pairs at 240 V bias. The systematics are dominated by our assumption about the 1.9 eV photon quantum yield. Lower quantum yield will result in a better baseline resolution. Based on this resolution and assuming that the leakage current leads to a flat shot-noise in the frequency domain and that our phonon pulses have a bandwidth of $\sim$3 kHz in the frequency domain, we can estimate a leakage current that is on the order of $10^{-16}$ A.  
This leakage current is almost an order of magnitude smaller than our earlier work using contact free geometry on a Ge detector with a similar volume, where the lowest baseline resolution achieved was $\sim 7$ eV ($\sim 2.4 e^{-}/h^{+}$ pair) \cite{Mirabolfathi, kadribasic_thesis}.

\section{\label{sec:sec4}Conclusion and outlook}

This paper reports a novel large mass Si phonon mediated solid state detector that can be used for Dark Matter and CE$\nu$NS experiments. The detector shows a baseline resolution of $\sim 1 e^{-}/h^{+}$ pair assuming a quantum yield of $\approx 1030$ $e^{-}/h^{+}$ pairs per laser pulse and a leakage current on the order of $10^{-16}$ A. This performance was achieved by implementing a contact free interface architecture and by removing an amorphous-Si layer. These features delayed the onset of the leakage current contribution to the noise with bias voltage while maintaining the linearity of the signal gain. This is a significant improvement over our earlier work with a Ge detector using contact free geometry, where the lowest baseline resolution achieved was $\sim 7$ eV ($\sim 2.4 e^{-}/h^{+}$ pair) \cite{Mirabolfathi} corresponding to a leakage current of $10^{-14}$ A. The use of pulsed laser photons as the source also allowed us to have a better control over the energy to avoid TES readout non-linearity. Such a low baseline resolution detector is already in use at the MINER experiment \cite{MINER}. Sensitivity to single $e^{-}/h^{+}$ pair excitation offers unique opportunities for experiments seeking signal via nuclear recoil interactions including the possibility of background discrimination in directional detection searches \cite{Kadribasic:2017obi}. As a future scope of work, we plan to use different laser photon energies to understand the quantum yield of Si and Ge.

\section*{Acknowledgements}
This work was supported by DOE grants de-sc0017859, de-sc0018981 and NSF grant NSF OISE 1743790. We acknowledge the contribution of the key cryogenic infrastructure (Bluefors LD400) provided by NISER, India. We would also like to acknowledge the support of DAE through the project Research in Basic Sciences - Dark Matter and SERB-DST through the J.C. Bose fellowship. The authors are very thankful to the CDMS UC Berkeley group who shared their phonon readout analysis and processing code package. The authors are also very thankful to Mr. Jason Caswell whose assistance was instrumental in building the experimental setup. We would also like to thank Dr. Kartikeswar Senapati, Dr . Ranbir Singh and Mr. Sourav Kundu for helpful discussions.

\bibliography{References}

\end{document}